\documentclass[prd,nofootinbib,superscriptaddress,twocolumn]{revtex4}

\usepackage{amsfonts,amssymb,amsthm,bbm}

\usepackage{amsmath}

\usepackage{hyperref}

\usepackage{color,psfrag}
\usepackage[dvips]{graphicx}

\usepackage{tikz}
\usetikzlibrary{calc}
\usetikzlibrary{decorations.pathmorphing}
\usetikzlibrary{shapes.geometric}
\usetikzlibrary{arrows,decorations.markings}

\newcommand{\C}{{\mathbb C}}

\newcommand{\R}{{\mathbb R}}

\newcommand{\cK}{{\mathcal K}}

\newcommand{\cH}{{\mathcal H}}

\newcommand{\cO}{{\mathcal O}}

\newcommand{\cC}{{\mathcal C}}
\newcommand{\cS}{{\mathcal S}}

\newcommand{\SU}{\mathrm{SU}}

\newcommand{\SL}{\mathrm{SL}}
\newcommand{\SO}{\mathrm{SO}}

\newcommand{\be}{\begin{equation}}
\newcommand{\ee}{\end{equation}}
\newcommand{\beq}{\begin{eqnarray}}
\newcommand{\eeq}{\end{eqnarray}}
\newcommand{\bes}{\begin{eqnarray}}
\newcommand{\ees}{\end{eqnarray}}

\newcommand{\mat} [2] {\left ( \begin{array}{#1}#2\end{array} \right ) }

\newcommand{\su}{{\mathfrak{su}}}

\renewcommand{\sl}{{\mathfrak{sl}}}
\newcommand{\so}{{\mathfrak{so}}}

\newcommand{\f}{\frac}

\def\nn{\nonumber}
\def\pp{\partial}

\def\rd{\mathrm{d}}

\def\ka{\kappa}

\def\eps{\epsilon}

\def\act{\triangleright}

\def\bw{\bar{w}}
\def\bW{\bar{W}}
\def\ttau{\tilde{\tau}}
\def\tx{\tilde{x}}
\def\ty{\tilde{y}}
\def\tX{\tilde{X}}
\def\tY{\tilde{Y}}

\def\vcK{\vec{\cK}}
\def\vvsigma{\vec{\varsigma}}
\def\veta{\vec{\eta}}

\def\rew{\mathfrak{Re}(w)}
\def\reW{\mathfrak{Re}(W)}
\def\imw{\mathfrak{Im}(w)}
\def\imW{\mathfrak{Im}(W)}

\begin{document}

\title{The Cosmological Spinor}

\author{{\bf Jibril Ben Achour}}
\affiliation{Center for Gravitational Physics, Yukawa Institute for Theoretical Physics, Kyoto University, 606-8502, Kyoto, Japan}
\author{{\bf Etera R. Livine}}\email{etera.livine@ens-lyon.fr}
\affiliation{Universit\'e de Lyon, ENS de Lyon, CNRS, Laboratoire de Physique LPENSL, 69007 Lyon, France}

\date{\today}

\begin{abstract}

We build upon previous investigation of the one-dimensional conformal symmetry of the Friedman-Lema\^ itre-Robertson-Walker (FLRW) cosmology of a free scalar field  and make it explicit through a reformulation of the theory at the classical level in terms of a manifestly $\textrm{SL}(2,\mathbb{R})$-invariant action principle. The new tool is a canonical transformation of the cosmological phase space to write it in terms of a spinor, i.e. a pair of complex variables that transform under the fundamental representation of $\textrm{SU}(1,1)\sim\textrm{SL}(2,\mathbb{R})$. The resulting FLRW Hamiltonian constraint is simply quadratic in the spinor and  FLRW cosmology is written as a Schr\"odinger-like action principle. Conformal transformations can then be written as proper-time dependent $\textrm{SL}(2,\mathbb{R})$ transformations. We conclude with possible generalizations of FLRW to arbitrary quadratic Hamiltonian and discuss the interpretation of the spinor as a gravitationally-dressed matter field or matter-dressed geometry observable.

\end{abstract}

\maketitle

\section*{Introduction}

This short paper explores the hidden conformal symmetry of the Friedman-Lema\^itre-Robertson-Walker (FLRW) cosmology of a free scalar field uncovered in \cite{BenAchour:2019ufa}. We aim at writing the FLRW cosmological action in the simplest way possible that makes this conformal invariance explicit.

Cosmology is a -if not the- natural arena of application of general relativity and of quantum gravity investigation, both due to its symmetries simplifying the theoretical analysis and to its physical relevance as a description of the evolution of the universe with measurable predictions.
Here we study FLRW cosmology, with general relativity coupled to a homogeneous and isotropic free scalar field, and focus on the background dynamics without inhomogeneities or other matter fields. This provides the simplest possible gravitational theory, which can serve as a toy model for illustrating the symmetries of general relativity and a test bed for quantum gravity methods. It can nevertheless hold surprises and its hidden conformal invariance was recently put forward in \cite{BenAchour:2019ufa}.
This conformal symmetry came out of earlier work introducing the CVH algebra of cosmological observables \cite{BenAchour:2017qpb,BenAchour:2018jwq,BenAchour:2019ywl}: the Poisson brackets of the geometry dilatation generator (or complexifier), of the spatial volume and of the Hamiltonian constraint form a closed $\sl(2,\R)$ Lie algebra, which can be carried to the quantum level and  fixes the operator ordering of the Wheeler-de Witt equation.
It was then understood that this CVH algebra plays a more fundamental role, as the Noether charges for a one-dimensional conformal invariance of FLRW cosmology under $\SL(2,\R)$ transformations acting as Mobius transformations in proper time \cite{BenAchour:2019ufa}. Moreover it allows one to map the FLRW cosmology of the free scalar field onto de Alfaro, Fubini and Forlan's conformal quantum mechanics \cite{deAlfaro:1976vlx}. As such, it becomes essential to preserve it during quantization or to provide a possible anomaly with a physical meaning.
Moreover, preserving this $\sl(2,\R)$ structure or not provides a strong criteria to classify possible regularizations of FLRW cosmology and modified gravity theories \cite{BenAchour:2018jwq,BenAchour:2019ywl}.
This structure holds independently of the value or sign of the cosmological constant \cite{ConformalLambda}.
Finally, it was shown that the CVH algebra can be embedded into a larger $\so(3,2)$ algebra of observables representing the whole phase space of homogeneous geometry and scalar field, and that this $\so(3,2)$ Lie algebra can be constructed from a $\su(1,1)\sim\sl(2,\R)$ spinor   which can be used as canonical variable for the quantization \cite{BenAchour:2020njq}.
The present work presents a last chapter to this story before embarking in the wider project of applying this conformal symmetry to the dynamics of inhomogeneities and extending it to midi-superspace models and full general relativity. Concrete applications of the conformal symmetry to the thermodynamics of FLRW cosmology will also be presented later on. Therefore, this work aims at writing explicitly the spinor, representing the whole phase space of FLRW cosmology coupled to a scalar field, and thereby making the conformal invariance appear limpidly.

\medskip

The first section deals with the classical reformulation of FLRW cosmology in terms of a spinor and a quadratic Hamiltonian.
Starting from the reduced Einstein-Hilbert action for general relativity coupled to an isotropic and homogeneous scalar field, we review its Hamiltonian formulation in terms of a four-dimensional phase space parametrized by the spatial volume $v$, its conjugate momentum (extrinsic curvature) $b$, the scalar field $\phi$ and its conjugate momentum $\pi_{\phi}$. We introduce the canonical transformation from those two pairs of canonical variables, $(v,b)$ representing the geometry and $(\phi,\pi_{\phi})$ representing the scalar matter, to a pair of complex variables $(w,W,\bw,\bW)$ mixing the geometry and scalar field.
First, we show that the Hamiltonian constraint $\cH$  is simply quadratic in the the complex variables, being the product of the imaginary parts of $w$ and $W$. This allows to write  FLRW cosmology as a Schr\"odinger action principle for the complex vector $(w,W)$, whose equation of motion is a Schr\"odinger equation straightforward to integrate. This is the simplest reformulation of the  FLRW background dynamics.
Second, we show that $w$ and $W$ live in the fundamental representation of the $\SU(1,1)$ group action generated by the CVH algebra of observables
formed by the Hamiltonian constraint $\cH$, the spatial volume $v$ and the integrated extrinsic curvature $\cC=vb$. This leads us to refer to $(w,W)$ as  the {\it cosmological spinor}.
More generally, the space of all quadratic polynomials in the spinor form the $\so(3,2)$ algebra of cosmological observables noticed in \cite{BenAchour:2020njq}, which can serve a basis for the quantization of the theory.

The second section deals with the symmetry of the theory. Written in term of the spinor $(w,W)$ and a quadratic Hamiltonian constraint dictating the evolution in proper time, the theory has a manifest $\SL(2,\R)\sim\SU(1,1)$ symmetry. We prove that combining the $\SL(2,\R)$ action with Mobius transformations in proper time leads to the symmetry of FLRW cosmology under one-dimensional conformal  transformations presented in \cite{BenAchour:2019ufa}. Finally, we show how to extend this action to  conformal transformations associated to arbitrary $\textrm{Diff}(\cS_{1})$ transformations of the proper time. These are not direct symmetries of the theory but the resulting action variation is a very interesting Schwarzian action, as explained in \cite{ConformalLambda}.

\section{FLRW Cosmology as an Evolving Spinor}

We consider a massless scalar field $\phi$ minimally coupled to gravity. 
The flat Friedman-Lema\^\i tre-Robertson-Walker (FLRW) mini-superspace model is defined by studying four-dimensional homogeneous isotropic metrics foliated by three-dimensional flat slices,
\be
\rd s^2 =-N(t)^2 \rd t^2 +a(t)^2\delta_{ij}\rd x^i\rd x^j
\,,
\ee
in terms of the lapse function $N(t)$ and the scalar factor $a(t)$. We similarly assume that the scalar field $\phi$ is homogeneous and only depends on the time $t$.
The reduced FLRW cosmological action, with vanishing cosmological constant $\Lambda=0$,  is given by the integration of the Einstein-Hilbert action over a fiducial three-dimensional cell of volume $V_{o}$ and reads:
\be
\label{FRWaction}
S[a,N,\phi]
\,=\,
V_{o}\int \rd t\left[
-\f{3}{8\pi G}\f{a\dot{a}^2}{N}+\f{a^3}{2N}\dot\phi^2
\right]
\,,
\ee
with the Newton gravitational constant $G$.
The canonical analysis of this action defines the conjugate momenta to the scale factor and scalar field:
\be
\pi_{a}=-\f{3V_{o}}{4\pi G}\f{a\dot a}{N}
\,,\qquad
\pi_{\phi}=\f{a^3V_{o}}{N}\dot\phi
\,,
\ee
and writes the action as a fully constrained system,
\be
S[a,N,\phi]=\int\rd t\,\big{[}
\dot a \pi_{a}+\dot \phi \pi_{\phi}-N\cH
\big{]}
\,.
\ee
The lapse $N$ plays the role of a Lagrange multiplier and the Hamiltonian constraint $\cH$ is a balance equation between the energy of the matter field and the energy of the geometry:
\be
\cH=\f1{2V_{o}}\left(
\f{\pi_{\phi}^2}{a^3}-\f{4\pi G}3 \f{\pi_{a}^2}a
\right)
\,.
\ee
It is convenient to introduce a volume variable, absorbing the volume of the fiducial cell. The  canonical transformation reads:
\be
v=a^3 V_{o}
\,,\qquad
b=-\f1{3V_{o}}\f{\pi_{a}}{a^2}=\f{1}{4\pi G}\f{\dot a}{Na}
\,.
\ee
Then the phase space of  FLRW cosmology is given by canonical variables,
\be
\{ b, v \} = 1
\,,\qquad
\{ \phi, \pi_{\phi}\} = 1
\,.
\ee
The Hamiltonian constraint becomes:
\beq
\cH
&=&
\f12\left(
\f{\pi_{\phi}^2}v-\ka^2 vb^2
\right)
\\
&=&
\f1{2v}
(\pi_{\phi}-\ka bv)(\pi_{\phi}+\ka bv)
\,\simeq0\,,
\nn
\eeq
where the the Planck length $\ka=\sqrt{12\pi G}$ (up to a numerical factor)  encodes the coupling between matter and geometry.
The FLRW action then reads
\be
S[\phi,\pi_{\phi},b,v,N]=\int\rd t\,\big{[}
\dot b v+\dot \phi \pi_{\phi}-N\cH
\big{]}
\,,
\ee
where we can switch kinetic term for the geometry to $-\dot v b$ by a total derivative. We can also write it  in terms of proper time $\tau$ defined by $\rd\tau =N\rd t$, absorbing the lapse into the time coordinate:
\be
S[\phi,\pi_{\phi},b,v,N]=\int\rd \tau\,
\Big{[}
v\rd_{\tau} b +\pi_{\phi}\rd_{\tau} \phi -\cH
\Big{]}
\,,
\ee
where we shall never forget about the implicit role of the lapse imposing that the overall energy  vanishes, $\cH=0$.

\subsection{Canonical map to a spinor phase space}

The previous work \cite{BenAchour:2020njq} identified a $\so(3,2)$ algebra of observables extending the $\sl(2,\R)$ Lie algebra formed by the three cosmological observables, the dilatation generator $\cC=vb$, the 3D volume $v$ and the Hamiltonian constraint $\cH$ (called CVH for short). The cosmological evolution formulated as a $\SL(2,\R)$ flow at the classical level, and then the quantization in terms of $\sl(2,\R)$ and $\so(3,2)$ representations, both hinted to a parametrization of the cosmological phase space in terms of canonical complex vectors living in the fundamental representation of $\SU(1,1)\sim\SL(2,\R)$. However the precise expression of those complex variables were not explicitly given. Here we remedy this shortcoming and introduce the pair of dimensionless complex variables, defined assuming that the volume remains positive $v>0$:
\be
\label{defw}
w=\f1{\sqrt{2}}e^{-\f\ka2 \phi}\,\left[
\sqrt{\f{v}{\lambda\ka^{3}}}
+i\sqrt{\f{\lambda\ka^{3}}{v}}
\left(\f{\pi_{\phi}}\ka-bv\right)
\right]
\,,
\ee
\be
\label{defW}
W=\f1{\sqrt{2}}e^{+\f\ka2 \phi}\,\left[
\sqrt{\f{v}{\lambda\ka^{3}}}
-i\sqrt{\f{\lambda\ka^{3}}{v}}
\left(\f{\pi_{\phi}}\ka+bv\right)
\right]
\,,
\ee
where $\lambda$ is an arbitrary dimensionless parameter defining the elementary unit of volume as $\lambda\ka^3$ in terms of the cubed Planck length.
Decomposing these variables in real and imaginary parts,
$w=x+iy$ and $W=X+iY$,
the spatial volume is  the product of the real parts:
\be
xX=\f{v}{2\lambda\ka^{3}}
\,,
\ee
while the Hamiltonian constraint is the product of the imaginary parts:
\be
yY
=
-\f{\lambda\ka}{2v}(\pi_{\phi}-\ka bv)(\pi_{\phi}+\ka bv)
=
-\lambda\ka\cH
\,.
\ee
The key point is that these new complex variables realize a canonical transformation of the four-dimensional phase space $(\phi,\pi_{\phi},b,v)$ and satisfy canonical Poisson brackets,
\beq
&&\{w,\bW\}=\{W,\bw\}=-i
\,,
\\
&&\{w,\bw\}=\{W,\bW\}=\{w,W\}=\{\bw,\bW\}=0
\,,\nn
\eeq
%
Indeed a lengthy though straightforward calculation for the kinetic term yields:
\be
\bW\rd w -W \rd \bw+\bw\rd W - w\rd\bW
=
-2i\big{[}
\pi_{\phi}\rd\phi+v\rd b
\big{]}\,,
\ee
so that the FLRW action can be simply re-written  as:
\beq
S&=\int \rd\tau\,
\bigg{[}&
\f i2\big{(}\bW\rd_{\tau} w -W \rd_{\tau} \bw+\bw\rd_{\tau} W - w\rd_{\tau}\bW\big{)}
\nn\\
&&-\f1{4\lambda\ka}(w-\bw)(W-\bW)
\,\bigg{]}
\,,
\label{wWaction}
\eeq
on in terms of the real and imaginary parts:
\be
S=\int \rd\tau\,
\bigg{[}
\big{(}Y\rd_{\tau} x-X\rd_{\tau} y +y\rd_{\tau} X-x\rd_{\tau} Y \big{)}
+\f1{\lambda\ka}yY
\bigg{]}
\,.
\nn
\ee
The big advantage of this reformulation is that the introduction of the variables $(w,W)\in\C^2$, which subtly mix the geometry with the scalar matter field, allows to write the theory with a simple quadratic Hamiltonian. As we will see later, this allows to think of the classical FLRW cosmology as a  Schr\"odinger action principle and to integrate the equation of motion as a mere exponentiation of a matrix Hamiltonian. 

Finally, we will refer to the variables $(w,W)$ as a spinor since they live in the fundamental representation of the $\SU(1,1)$ transformation generated by the CVH observables, as shown below.

\subsection{Cosmological evolution and CVH algebra}

The quadratic Hamiltonian $\cH$ gives the evolution in proper time, $\rd_{\tau}\cO=\{\cO,\cH\}$ for any observable $\cO$. This gives the equation of motion for the spinor:
\beq
&&\cH=\f1{4\lambda\ka}(w-\bw)(W-\bW)
\nn\\
&&
\Rightarrow\quad
\rd_{\tau}w
=\rd_{\tau}\bw
=\f i{4\lambda\ka}(w-\bw)\,,
\eeq
and similarly  for the second complex variable:
\be
\rd_{\tau}W
=
\rd_{\tau}\bW
=
\f i{4\lambda\ka}(W-\bW)
\,.
\ee
The imaginary parts of $w$ and $W$ obviously are both  constants of motion:
\be
\{\cH,w-\bw\}=\{\cH,W-\bW\}=0\,.
\ee
These Dirac observables allow to integrate the evolution for  the spinor:
\be
e^{-\{\tau H,\cdot\}}\,w=w+\tau\{w,\cH\}=
w+\f{i\tau}{4\lambda\ka}(w-\bw)
\,,
\ee
and similarly for $W$.
This linear evolution is easily translated in terms of  real and imaginary parts:
\be
\left|
\begin{array}{lcl}
y(\tau)&=&y_{0}\vspace*{1mm}\\
x(\tau)&=&x_{0}-\f{\tau}{2\lambda\ka}y_{0}
\end{array}
\right.
\ee

As shown in \cite{BenAchour:2017qpb,BenAchour:2019ufa}, the dynamics can also be neatly re-packaged in terms of the CVH observables. These observables are here realized as quadratic polynomials in the spinor:
\be
\left|\begin{array}{lcl}
\cH&=&\f1{4\lambda\ka}(w-\bw)(W-\bW)
\,,\vspace*{2mm}\\
v&=&\f{\lambda\ka^{3}}2(w+\bw)(W+\bW)
\,,\vspace*{2mm}\\
C&=&\f1{2i}(\bw\bW-wW)
\,.
\end{array}\right.
\ee
Indeed, this set of observables directly gives  the evolution equations for the volume:
\be
\left|\begin{array}{lcl}
\rd_{\tau}v&=&\{v,\cH\}=\ka^{2}C
\,,\vspace*{1mm}\\
\rd_{\tau}C&=&\{C,\cH\}=-\cH
\,,\vspace*{1mm}\\
\rd_{\tau}\cH&=&0\,.
\end{array}\right.
\ee
Completed by the Poisson bracket $\{C,v\}=v$, these Poisson brackets form a closed $\sl(2,\R)$ Lie algebra, referred to unimaginatively as the CVH algebra. More precisely, we can write the observables $C$, $v$ and $\cH$ in terms of the usual $\sl(2,\R)$ basis $(j_{z},k_{x},k_{y})$:
\be
 C = k_y 
 \,,\,\,
 v = \lambda\ka^{3}( k_x + j_z )
 \,,\,\,
 \cH=  \frac{1}{2\lambda\ka} \big{(} k_x - j_z \big{)}
 \,,
\ee
\be
\left|\begin{array}{lcl}
j_z
&=&
\f v{2\lambda\ka^{3}}-\lambda\ka\cH
=
\f12\big{(}
\bw W+w \bW
\big{)}
\,,\vspace*{1mm}\\
k_x &=&
\f v{2\lambda\ka^{3}}+\lambda\ka\cH
=
\f12\big{(}
w W+\bw \bW
\big{)}
\,,\vspace*{1mm}\\
k_y &=& \f1{2i}\big{(}\bw\bW-wW\big{)}
\,,
\end{array}\right.
\ee
These quadratic observables $\vcK=(j_{z},k_{x},k_{y})$ actually generate $\SU(1,1)$ transformations on the variables $w$ and $W$:
\be
\label{SU11action}
\left\{\veta\cdot\vcK,\mat{c}{w\\\bw}\right\}
\,=\,
\f i2\veta\cdot\vvsigma\,\mat{c}{w\\\bw}
\,,
\ee
and similarly for $W$, with the  Lorentzian signature scalar product $\veta\cdot\vcK=\eta_{z}j_{z}-\eta_{x}k_{x}-\eta_{y}k_{y}$,
where the matrices $\varsigma_{i}$ are the Lorentzian Pauli matrices,
\be
\varsigma_{z}=\mat{cc}{1 & 0 \\ 0 & -1}
\,,\quad
\varsigma_{x}=\mat{cc}{ 0 & 1 \\ -1 &0}
\,,\quad
\varsigma_{y}=\mat{cc}{ 0 & -i \\-i &0}
\,,
\nn
\ee
\be
[\varsigma_{z},\varsigma_{x}]=2i\varsigma_{y}
\,,\quad
{[}\varsigma_{x},\varsigma_{y}]=-2i\varsigma_{z}
\,,\quad
{[}\varsigma_{y},\varsigma_{z}]=2i\varsigma_{x}
\,.
\ee
This justifies to refer to $(w,W)$ as a spinor.
In particular, as noticed in \cite{BenAchour:2017qpb}, the Hamiltonian constraint is the null generator $\cH=  ( k_x - j_z )/2\lambda\ka$, whose action can be exponentiated in terms of $\SU(1,1)$ matrices:
\be
e^{\tau\{\cH,\cdot\}}\,\mat{c}{w\\\bw}
=
e^{i\f \tau{4\lambda \ka}(\varsigma_{x}-\varsigma_{z})}
\,\mat{c}{w\\\bw}\,,
\ee
which leads back to the trajectory for $w$ derived earlier.

\medskip

More generally, all quadratic polynomial in $w$ and $W$ form a $\so(3,2)$ Lie algebra of observables \cite{BenAchour:2020njq}. Indeed, the ten quadratic combinations of $w$, $W$ and their complex conjugate can be combined into the ten generators of $\so(3,2)$:
\be
\begin{array}{lll}
J_{z}=\f12(w\bW+\bw W)
\,,&
J_{-}=\f12(W^2-w^2)
\,,&
J_{+}=\overline{J_{-}}
\,,\vspace*{1mm}\\
K_{z}=\f12(w\bw- W\bW)
\,,&
K_{-}=\f12wW
\,,&
K_{+}=\overline{K_{-}}
\,,\vspace*{1mm}\\
L_{z}=\f i2(\bw W-w \bW)
\,,&
L_{-}=\f i2(w^2+W^2)
\,,&
L_{+}=\overline{L_{-}}
\,,\vspace*{1mm}\\
D=\f12(w\bw+ W\bW)&&
\end{array}
\nn
\ee
whose Poisson brackets are:
\be
\begin{array}{l}
\{D,J_{a}\}=0
\,,\quad
\{D,K_{a}\}=L_{a}
\,,\quad
\{D,L_{a}\}=-K_{a}
\,,\vspace*{1mm}\\
\{J_{z},J_{\pm}\}=\mp i J_{\pm}
\,,\,\,
\{J_{z},K_{\pm}\}=\mp i K_{\pm}
\,,\,\,
\{J_{z},L_{\pm}\}=\mp i L_{\pm}
\,,
\vspace*{1mm}\\
\{K_{z},J_{\pm}\}=\mp i K_{\pm}
\,,\quad
\{L_{z},J_{\pm}\}=\mp i L_{\pm}
\,,\vspace*{1mm}\\
\{L_{z},L_{\pm}\}=\pm i J_{\pm}
\,,\quad
\{K_{z},K_{\pm}\}=\pm i J_{\pm}
\,,\vspace*{1mm}\\
\{K_{z},L_{\pm}\}=
\{L_{z},K_{\pm}\}=
\{J_{z},K_{z}\}=
\{J_{z},L_{z}\}=
0
\,,\vspace*{1mm}\\
\{J_{+},J_{-}\}=-2iJ_{z}
\,,\quad
\{K_{+},K_{-}\}=\{L_{+},L_{-}\}=2i J_{z}
\,,\vspace*{1mm}\\
\{J_{-},K_{+}\}=\{K_{-},J_{+}\}=2iK_{z}
\,,\quad
\{J_{+},K_{+}\}=0
\,,\vspace*{1mm}\\
\{J_{-},L_{+}\}=\{L_{-},J_{+}\}=2iL_{z}
\,,\quad
\{J_{+},L_{+}\}=0
\,,\vspace*{1mm}\\
\{K_{z},L_{z}\}=-D
\,,\quad
\{K_{+},L_{-}\}=-2D
\,,\quad
\{K_{+},L_{+}\}=0
\,,
\end{array}
\nn
\ee
These $\so(3,2)$ generators can be translated into cosmological observables by expressing the spinors back in terms of the volume and scalar field. The $\so(3,2)$ algebra of observables is much larger than the CVH algebra. Preserving the $\so(3,2)$ structure becomes especially relevant during the quantization process \cite{BenAchour:2020njq}, but it is not directly relevant to the present work.
Nevertheless, among these  $\so(3,2)$ generators, we  identify a Casimir of the CVH algebra as:
\be
\{\bw W-w\bW,\cO\}=0
\quad\textrm{for}\quad\cO=\cH,v,C\,.
\ee
Not only can it be understood as the square-root of the quadratic Casimir $(k_{x}^2+k_{y}^2-j_{z}^2)$ of the $\sl(2,\R)$ Lie algebra, but it  also turns out to simply be the scalar field momentum once translated back into cosmological quantities using the definition \eqref{defw}-\eqref{defW} of the spinor:
\be
\f i 2(\bw W-w\bW)
=
\f{\pi_{\phi}}\ka
\,.
\ee

\subsection{Classical cosmology as a Schr\"odinger action principle}

In light of the algebraic structure described above, it seems natural to organize the complex variables as a 4-dimensional complex vectors
$\Phi=(w,\,\bw,\,W,\,\bW)$.
Then the FLRW action \eqref{wWaction} can be written as:
\be
S[\Phi]=
\int\rd\tau\,\bigg{[}
\f i2\Phi^*\rd_{\tau}\Phi + \f1{8\lambda\ka}\Phi^*H_{eff} \Phi
\bigg{]}
\,,
\ee
\be
\textrm{with}\quad
H_{eff}=\mat{c|c}{\sigma & 0 \\ \hline 0 & \sigma}
\,,\quad
\sigma=\varsigma_z-\varsigma_x
\,.\nn
\ee
We have introduced a slightly modified complex conjugation:
\be
\Phi^*=(\gamma\Phi)^\dagger=\Phi^\dagger\gamma\,,
\quad\textrm{with}\quad
\gamma=\mat{c|c}{0 & \varsigma_z \\ \hline \varsigma_z & 0}
\,.
\ee
%
This is a Schr\"odinger action principle whose equation of motion is a Schr\"odinger-like equation on the spinor $\Phi$:
\be
\rd_{\tau}\Phi=\f{i}{4\lambda\ka}H_{eff}\Phi\,.
\ee
Since the effective Hamiltonian $H_{eff}/4\lambda\ka$ is diagonal by block, the evolution of the two variables $w$ and $W$ decouples:
\be
\pp_{\tau}\mat{c}{w\\\bw}=\f{i}{4\lambda\ka} \sigma \mat{c}{w\\\bw}
\,,\quad
\pp_{\tau}\mat{c}{W\\\bW}=\f{i}{4\lambda\ka} \sigma \mat{c}{W\\\bW}
\,,
\nn
\ee
with the cosmological trajectory given by the exponentiation of the Hamiltonian:
\be
\mat{c}{w(\tau)\\\bw(\tau)}
=
e^{\f{i\,\tau}{4\lambda\ka} \sigma}
\,\mat{c}{w_{0}\\\bw_{0}}
\,,
\ee
and similarly for $W(\tau)$.
This exactly matches the FLRW Hamiltonian derived from the CVH algebra as described above as the $(k_{z}-j_{x})$ generator of the $\su(1,1)\sim\sl(2,\R)$ Lie algebra.

Let us not forget that physical cosmological trajectories must satisfy the Hamiltonian constraint, i.e. that the imaginary part of $w$ or of $W$ vanishes. As one can see form the equations above, the imaginary part, $(w-\bw)$ and $(W-\bW)$, are conserved during the evolution. So it is enough to require that one of them vanishes as initial condition.

\section{Conformal Invariance}

While the CVH $\sl(2,\R)$ algebra for FLRW cosmology was first used to fix quantization and regularization ambiguities \cite{BenAchour:2017qpb,BenAchour:2018jwq,BenAchour:2019ywl}, it was only later understood  as a symmetry of the theory. in terms of conformal symmetry. Indeed, in \cite{BenAchour:2019ufa}, it was shown that FLRW cosmology is invariant under 1D conformal transformations realized  as Mobius transformations in proper time and the associated Noether charges were identified as the initial conditions of the CVH observables.
However the $\SL(2,\R)\sim\SU(1,1)$ transformations generated by exponentiating the Poisson brackets with the CVH observables, as derived above in \eqref{SU11action}, are not symmetry of the theory. The obvious difference with the 1D conformal transformation is that the $\SL(2,\R)$ transformations are time-independent while the Mobius transformations are clearly time-dependent.
We revisit this here and clarify the action of conformal transformations on the cosmological phase space. In particular, we realize them as time-dependent $\SL(2,\R)$ transformations on the spinors.

\subsection{$\SL(2,\R)$ transformations as symmetries}

Starting with the FLRW action written in terms of the real and imaginary parts of the spinor,
\be
S=\int \rd\tau\,
\bigg{[}
\big{(} Y\rd_{\tau} x-X\rd_{\tau} y +y\rd_{\tau} X -x\rd_{\tau} Y \big{)}
+\f1{\lambda\ka}yY
\bigg{]}
\,,
\nn
\ee
it is convenient to re-write it in terms of real 2-vectors in order to come back to a $\SL(2,\R)$ action:
\be
\psi=\mat{c}{x \\ y}
\,,\qquad
\Psi=\mat{c}{X \\ Y}
\,.
\ee
Indeed, the $\SU(1,1)$ action on the spinor, $G(w,\,\bw)$ and $G(W,\,\bW)$, with the complex matrix
\be
G=\mat{cc}{a & b \\ \bar{a} & \bar{b}} \quad\textrm{with}\quad |a|^{2}-|b|^{2}=1
\,,\nn
\ee
translates into the $\SL(2,\R)$ action on the spinor's real and imaginary parts, $M(x,\,,y)$ and $M(X,\,Y)$ with the  matrix
\be
M=\mat{cc}{\alpha & \beta \\ \gamma & \delta} \quad\textrm{with}\quad \alpha\delta-\beta\gamma=1
\,,\nn
\ee
whose real components $\alpha$, $\beta$, $\gamma$, $\delta$ are given by $\mathfrak{Re}(a)\pm\mathfrak{Re}(b)$ and  $\mathfrak{Im}(a)\pm\mathfrak{Im}(b)$.

\smallskip

In terms of the real 2-vectors $\psi$ and $\Psi$, the FLRW action becomes:
\be
S=-\int\rd\tau
\Big{[}
{}^{t}\Psi \eps\rd_{\tau}\psi
+
{}^{t}\psi \eps\rd_{\tau}\Psi
-
\f1{\lambda\ka}\,{}^{t}\Psi P \psi
\Big{]}
\ee
\be
\textrm{with}\quad
\eps=\mat{cc}{ 0 & 1 \\ -1 & 0}
\quad\textrm{and}\quad
P=\mat{cc}{ 0 &0 \\ 0 & 1}
\,,\nn
\ee
with ${}^{t}\psi \eps\rd_{\tau}\Psi=-(\rd_{\tau}{}^{t}\Psi )\,\eps\psi$.
%
%
%
The free theory, defined by ignoring the Hamiltonian term and focusing on the kinetics term $({}^{t}\Psi \eps\rd_{\tau}\psi+{}^{t}\psi \eps\rd_{\tau}\Psi)$, is invariant under time independent $\SL(2,\R)$ transformation, $\psi\mapsto M\psi$ and $\Psi\mapsto M\Psi$. To extend this to a $\SL(2,\R)$ symmetry for the evolving theory, we  need to take into account the evolution dictated by the Hamiltonian term  ${}^{t}\Psi P \psi$.
We thus consider the time dependent $\SL(2,\R)$ transformations:
\be
\left|
\begin{array}{lcl}
\psi&\mapsto& M(\tau)\psi\,, \vspace*{2mm}\\
\Psi&\mapsto& M(\tau)\Psi \,,
\end{array}
\right.
\qquad\textrm{where}\quad 
M(\tau)= \cO_{\tau}^{{-1}}M \cO_{\tau}
\,,
\nn
\ee
\be
\textrm{with}\quad 
 \cO_{\tau}=e^{+\f{\tau}{2\lambda\ka}\eps P}=\mat{cc}{1 & \f{\tau}{2\lambda\ka} \\ 0 & 1}
\,.
\ee
Using that $\SL(2,\R)$ matrices preserves the symplectic canonical form, ${}^{t}M\eps M=\eps$, a simple calculation allows to check that the Lagrangian is indeed invariant under such modified $\SL(2,\R)$ transformations.

If we write explicitly those transformations,
these do not match the 1D conformal transformations introduced in \cite{BenAchour:2019ufa}. In order to identify these two symmetries, we need to combine the $\SL(2,\R)$ transformation of the spinor with a Mobius transformation of the proper time, as explained below.

\subsection{Conformal map on the spinor}

We combine the  time dependent $\SL(2,\R)$ transformation of the spinor with a transformation of the proper time and consider:
\be
M=\mat{cc}{\alpha & \beta \\ \gamma & \delta}\in\SL(2,\R)
\,,\quad
\left|
\begin{array}{lcl}
\psi&\mapsto&
\widetilde{\psi}=\cO_{\ttau}^{-1}M\cO_{\tau}\psi
\,,
\vspace*{2mm}\\
\Psi&\mapsto&
\widetilde{\Psi}=\cO_{\ttau}^{-1}M\cO_{\tau}\Psi\,,
\end{array}
\right.
\nn
\ee
\be
\textrm{with}\quad
\tau \mapsto \ttau=M\act \tau = \ttau=2\lambda\ka\,\f{\alpha\f{\tau}{2\lambda\ka}+\beta}{\gamma\f{\tau}{2\lambda\ka}+\delta}
\,,
\ee
This map consists in  $\SL(2,\R)$-transformations on the evolved spinor, $\cO_{\ttau}\widetilde{\psi}=M\,\cO_{\tau}\psi$.
$\ka$ factors were ignored in previous work, but keeping track of them allows to respect the physical dimension of $\tau$ as a time and the transformation parameters $(\alpha,\beta,\gamma,\delta)$ dimensionless.
Real and imaginary parts of the complex variables get rescales in opposite ways, plus an extra shift for the imaginary part:
\be
\label{mobius}
\left|
\begin{array}{lcl}
x&\mapsto& \tx(\ttau)=\left(\gamma\f{\tau}{2\lambda\ka}+\delta\right)^{-1}\,x(\tau)
\,,\vspace*{1mm}\\
y&\mapsto&\ty(\ttau)={\left(\gamma\f{\tau}{2\lambda\ka}+\delta\right)}y(\tau)+\gamma x(\tau)\,,
\end{array}
\right.
\ee
and the same transformation for the second spinor component $W=X+iY$. 

First, this indeed defines a group action, representing the $\SL(2,\R)$ product as 2$\times$2 matrix multiplication.
Second,
a direct computation allows to check that these transformations leave the action invariant:
\be
\widetilde{S}=
\int\rd\ttau\,
\bigg{[}
\big{( \tY\rd_{\ttau} \tx-}\tX\rd_{\ttau} \ty +\ty\rd_{\ttau} \tX -\tx\rd_{\ttau} \tY \big{)}
+\f1{\lambda\ka}\ty\tY
\bigg{]}
=
S\,,
\nn
\ee
without any total derivative term. These conformal transformations, realized as time dependent $\SL(2,\R)$ transformations, therefore define a symmetry of the theory. These fit with the 1D conformal symmetry transformations introduced in \cite{BenAchour:2019ufa} and provide their explicit action on the spinorial phase space.

\subsection{From $\SL(2,\R)$ to $\textrm{Diff}(\cS_{1})$ conformal transformations}

It is possible to extend the transformation law of the cosmological spinor under $\SL(2,\R)$ Mobius transformations in proper time to arbitrary mappings of the proper time. By mimicking the transformations given above, we propose that the real and imaginary parts of the complex variables transform under a map $\tau\mapsto\ttau=f(\tau)$ as:
\be
\label{conformal}
\left|
\begin{array}{lcl}
x&\mapsto& \tx(\ttau)=h^{1/2}{x(\tau)}
\,,\vspace*{1mm}\\
y&\mapsto&\ty(\ttau)
= h^{-1/2}y(\tau)+2\lambda\ka(\rd_{\tau}h^{-1/2}) x(\tau)
\end{array}
\right.
\,,
\ee
where $h=\rd_{\tau}f$ is the Jacobian of the transformation of the proper time. These define a group  action representing $\textrm{Diff}(\cS_{1})$. Moreover, a straightforward calculation allows to compute the resulting variation of the action:
\beq
\widetilde{S}-S
&=&2\lambda\ka\int \rd\tau\,\bigg{[}
h^{-1}\rd_{\tau}^2h-\f32h^{-2}(\rd_{\tau}h)^2
\bigg{]}
Xx\nn\\
&=&2\lambda\ka\int \rd\tau\,\textrm{Sch}[f]
Xx
\,,
\label{conformalvariation}
\eeq
without any total derivative term. This variation involves the Schwarzian derivative $\textrm{Sch}[f]$. We recognize the volume factor $v=2\lambda\ka^3xX$. This allows to recover the same conformal transformations as introduced in \cite{BenAchour:2019ufa,ConformalLambda}. When the Schwarzian vanishes, we recover the invariance of FLRW cosmology under the $\SL(2,\R)$ group of Mobius transformations. 
In the general case, the $(\lambda\ka)$ factor in the transformation law  \eqref{conformal} for the imaginary part of the spinor  makes it clear that this extra-term is directly related to the Hamiltonian of the theory $\cH=yY/\lambda\ka$.
Moreover, on top of deriving a Schwarzian action, one could wonder if extensions of the $\textrm{Diff}(\cS_{1})$ transformations \eqref{conformal} could create ex-terms in the action variation \eqref{conformalvariation} which would still be quadratic in the spinor. Generating such extra effective quadratic terms to the Hamiltonian constraint would explore the whole $\so(3,2)$ algebra of FLRW observables \cite{BenAchour:2020njq} and might lead to some physically-interesting corrections to  FLRW cosmology.

\section*{Outlook}

To summarize the work presented here, we introduced a spinorial reformulation of the FLRW cosmology coupling a scalar field to homogeneous and isotropic general relativity.
Starting from the four-dimensional phase space parametrized by the canonical variables, the spatial volume and extrinsic curvature $(v,b)$ for the geometry sector and the scalar field and momentum $(\phi,\pi_{\phi})$ for the matter sector, we defined a canonical transformation to a pair of complex variables $(w,W)\in\C^2$. The Hamiltonian constraint, generating the cosmological evolution in proper time, becomes simply the product of the imaginary parts $\imw\imW$, while the volume becomes the product of the real parts $\rew\reW$. The FLRW action then reads (up to a total derivative):
\be
\label{finalFLRWaction}
S_{FLRW}=\int \rd\tau\,
\bigg{[}
i\big{(}\bW\rd_{\tau} w -W \rd_{\tau} \bw{)}
+\f1{\lambda\ka}\imw\imW
\bigg{]}
\,,
\nn
\ee
where $\ka$ is the Planck length (up to a numerical factor) and $\lambda$ is an arbitrary real parameter entering the mapping between $(v,b,\phi,\pi_{\phi})$ and $(w,W)$. Here the lapse $N$ implicitly enters the definition of the proper time from the time coordinate $\rd\tau=N\rd t$ and enforces that the Hamiltonian vanishes on-shell, i.e. $\imw\imW=0$. The case $\imw=0$ corresponds the expanding phase of FLRW cosmology (with $\pi_{\phi}=+\ka bv$), while $\imW=0$ corresponds to the contracting phase (with $\pi_{\phi}=-\ka bv$).

This especially simple action with a quadratic Hamiltonian is perfectly suited for the quantization for the theory. However, the quantization is not the goal we have pursued here. Instead, we have focused on the symmetries of the theory. The spinorial action is obviously invariant under $\SU(1,1)\sim\SL(2,\R)$ transformations on $w$ and $W$. We showed how this $\SL(2,\R)$ action combines with Mobius transformations in proper time into a 1D conformal symmetry\footnotemark{} of  FLRW cosmology. 
\footnotetext{
As showed in \cite{BenAchour:2019ufa}, the three Noether charges associated to those conformal transformations are the Hamiltonian constraint $\cH$, the volume $v$ and the integrated intrinsic curvature $\cC=vb$, which form the CVH algebra introduced in \cite{BenAchour:2017qpb}.
}
This symmetry should allow to quantize FLRW cosmology as a one-dimensional conformal theory, as suggested in \cite{BenAchour:2019ufa}.
Finally, we showed how to extend these $\SL(2,\R)$ conformal transformations into conformal transformations associated to arbitrary diffeomorphisms of the proper time, $\tau\mapsto\ttau=f(\tau)$. These are not symmetries per se of the theory, but the resulting variation of the action gives an enticing Schwarzian action for $f$.

\medskip

This formalism  has natural possible extensions:
\begin{itemize}

\item One could consider more general quadratic Hamiltonians. For instance, a term in $\rew\reW \propto v$  introduces a cosmological constant, and terms such as $\rew^2\propto ve^{-\ka\phi}$ or $\reW^2\propto ve^{+\ka\phi}$ introduce inflationary potentials for cosmology. Interestingly all quadratic Hamiltonian can be seen as part of a $\so(3,2)$ Lie algebra of observables and the corresponding evolution can be integrated as a $\SO(3,2)$ flow \cite{BenAchour:2020njq}. Moreover, this $\so(3,2)$ structure can be preserved under quantization and FLRW quantum cosmology defined as a $\SO(3,2)$ irreducible representation.

\item More generally, a quadratic Hamiltonian is usually the starting point for perturbation theory, thus the natural next step would be to introduce higher order terms, defining non-trivial interaction terms for the geometry and matter field. As the quadratic Hamiltonian can be written as a Schr\"odinger action principle, higher order terms would lead to a non-linear Schr\"odinger equations\footnotemark{}, whose cosmological interpretation should be investigated.
\footnotetext{
Let us point out the relation of FLRW cosmology to cubic and quintic non-linear Schr\"odinger equations put forward in \cite{Lidsey:2013osa,Lidsey:2018byv}.
}

\item One could also wonder if regularizations or quantum gravity corrections of FLRW cosmology (e.g. see \cite{Bojowald:2015iga}), such as the polymer regularization (also know as loop quantum cosmology) \cite{Ashtekar:2011ni,Agullo:2016tjh,Bojowald:2020wuc} or string cosmology \cite{Brandenberger_2011},
also admit a spinorial representation with a polynomial Hamiltonian constraint. For instance, there exists a version of polymerized FLRW cosmology, which regularizes the initial singularity into a big bounce and preserves the $\sl(2,\R)$ structure of the CVH algebra \cite{BenAchour:2018jwq}, that should be writable in similar terms with spinor variables mixing geometry and matter fields.
The spinorial formulation also suggests a natural method to regularize the initial singularity: we can combine the contracting and expanding cosmological phases by slightly modifying the Hamiltonian constraint $\cH=\imw\imW$, for example by introducing an energy shift $\imw\imW\rightarrow \imw\imW+\eps$ or a non-polynomial extension $\imw\imW\rightarrow (\imw\imW+1/\imw\imW)$. It would be interesting to understand what such $F[\imw\imW]$ Hamiltonian constraints imply physically depending on the properties of the chosen function $F$.

\end{itemize}
It would of course be better to understand if any of those extensions or modifications to FLRW cosmology can be derived as the homogeneous sector or the coarse-graining of a covariant  modified gravity theory or quantum gravity model.

Finally, an interesting feature of our approach is the use of variables mixing the matter and the geometry. Indeed, in light of the equivalence between geometry and matter proposed by the Einstein equations, it seems natural to introduce dressed variables, which include some geometry degrees of freedom in the matter field variables and vice-versa. This is exactly the role of the spinor which we define: combine matter and geometry in order to simplify the expression of the Hamiltonian constraint for the coupled system.
In order to go further, we would need to extend this approach to inhomogeneities, if not to full general relativity. Beyond mini- and midi- superspace models, one possible path would be to build an inhomogeneous cosmology as a network (or lattice) of interacting FLRW cells (e.g. as attempted in \cite{WilsonEwing:2012bx}), each cell with their own (generalized) spinor variable and their own (extended) CVH algebra. Their degrees of freedom attached to each cell could represent the homogeneous geometry of the space(-time) region after coarse-grained and/or gauge-fixing or the algebra of surface charges living on the cell boundary (e.g. as in \cite{Freidel:2019ees}).
At the continuum level, one might look for a link with the spinorial representation of Jackiw-Teitelboim gravity  \cite{Wieland:2020ogk}  in light of the relation  between our approach and Schwarzian mechanics \cite{ConformalLambda}, on the one hand, and between Schwarzian mechanics and Jackiw-Teitelboim gravity on the other hand \cite{Maldacena:2016upp,Mertens_2018}.
And more generally, it would be enlightening if to understand if there is a link with the use of spinors in general relativity, for instance  for the positive energy theorem \cite{Witten:1981mf,Parker:1981uy,Wieland:2016dbc,Wieland:2017zkf}.


%
%



\bibliographystyle{bib-style}
\bibliography{QC}

\providecommand{\href}[2]{#2}\begingroup\raggedright\begin{thebibliography}{10}

\bibitem{BenAchour:2019ufa}
J.~Ben~Achour and E.~R. Livine, ``{Cosmology as a CFT$_1$},'' JHEP {\bf 12}
  (2019) 031,
\href{http://arXiv.org/abs/1909.13390}{{\texttt{arXiv:1909.13390}}}.

\bibitem{BenAchour:2017qpb}
J.~Ben~Achour and E.~R. Livine, ``{Thiemann complexifier in classical and
  quantum FLRW cosmology},'' Phys. Rev. {\bf D96} (2017), no.~6, 066025,
\href{http://arXiv.org/abs/1705.03772}{{\texttt{arXiv:1705.03772}}}.

\bibitem{BenAchour:2018jwq}
J.~Ben~Achour and E.~R. Livine, ``{Polymer Quantum Cosmology: Lifting
  quantization ambiguities using a $SL(2,\mathbb{R})$ conformal symmetry},''
  Phys. Rev. {\bf D99} (2019), no.~12, 126013,
\href{http://arXiv.org/abs/1806.09290}{{\texttt{arXiv:1806.09290}}}.

\bibitem{BenAchour:2019ywl}
J.~Ben~Achour and E.~R. Livine, ``{Protected $SL(2,\mathbb{R})$ Symmetry in
  Quantum Cosmology},'' JCAP {\bf 1909} (2019) 012,
\href{http://arXiv.org/abs/1904.06149}{{\texttt{arXiv:1904.06149}}}.

\bibitem{deAlfaro:1976vlx}
V.~de~Alfaro, S.~Fubini, and G.~Furlan, ``{Conformal Invariance in Quantum
  Mechanics},'' Nuovo Cim. {\bf A34} (1976)
569.

\bibitem{ConformalLambda}
J.~Ben~Achour and E.~R. Livine, ``{Generating the cosmological constant from a
  conformal transformation},''
  \href{http://arXiv.org/abs/2004.05841}{{\texttt{arXiv:2004.05841}}}.

\bibitem{BenAchour:2020njq}
J.~Ben~Achour and E.~R. Livine, ``{Conformal structure of FLRW Cosmology:
  Spinorial representation and the so(3,2) algebra of observables},''
\href{http://arXiv.org/abs/2001.11807}{{\texttt{arXiv:2001.11807}}}.

\bibitem{Lidsey:2013osa}
J.~E. Lidsey, ``{Scalar Field Cosmologies Hidden Within the Nonlinear
  Schrodinger Equation},''
\href{http://arXiv.org/abs/1309.7181}{{\texttt{arXiv:1309.7181}}}.

\bibitem{Lidsey:2018byv}
J.~E. Lidsey, ``{Inflationary Cosmology, Diffeomorphism Group of the Line and
  Virasoro Coadjoint Orbits},''
\href{http://arXiv.org/abs/1802.09186}{{\texttt{arXiv:1802.09186}}}.

\bibitem{Bojowald:2015iga}
M.~Bojowald, ``{Quantum cosmology: a review},'' Rept.\ Prog.\ Phys. {\bf 78}
  (2015) 023901,
  \href{http://arXiv.org/abs/1501.04899}{{\texttt{arXiv:1501.04899}}}.

\bibitem{Ashtekar:2011ni}
A.~Ashtekar and P.~Singh, ``{Loop Quantum Cosmology: A Status Report},''
  Class.\ Quant.\ Grav. {\bf 28} (2011) 213001,
  \href{http://arXiv.org/abs/1108.0893}{{\texttt{arXiv:1108.0893}}}.

\bibitem{Agullo:2016tjh}
I.~Agullo and P.~Singh, {\em {Loop Quantum Cosmology}}, pp.~183--240.
\newblock WSP, 2017.
\newblock \href{http://arXiv.org/abs/1612.01236}{{\texttt{arXiv:1612.01236}}}.

\bibitem{Bojowald:2020wuc}
M.~Bojowald, ``{Critical evaluation of common claims in loop quantum
  cosmology},'' Universe {\bf 6} (2020), no.~3, 36,
  \href{http://arXiv.org/abs/2002.05703}{{\texttt{arXiv:2002.05703}}}.

\bibitem{Brandenberger_2011}
R.~H. Brandenberger, ``String gas cosmology: progress and problems,'' Classical
  and Quantum Gravity {\bf 28} (Oct, 2011) 204005.

\bibitem{WilsonEwing:2012bx}
E.~Wilson-Ewing, ``{Lattice loop quantum cosmology: scalar perturbations},''
  Class.\ Quant.\ Grav. {\bf 29} (2012) 215013,
  \href{http://arXiv.org/abs/1205.3370}{{\texttt{arXiv:1205.3370}}}.

\bibitem{Freidel:2019ees}
L.~Freidel, E.~R. Livine, and D.~Pranzetti, ``{Gravitational edge modes: from
  Kac--Moody charges to Poincar\'e networks},'' Class.\ Quant.\ Grav. {\bf 36}
  (2019), no.~19, 195014,
  \href{http://arXiv.org/abs/1906.07876}{{\texttt{arXiv:1906.07876}}}.

\bibitem{Wieland:2020ogk}
W.~Wieland, ``{Twistor representation of Jackiw-Teitelboim gravity},''
  \href{http://arXiv.org/abs/2003.13887}{{\texttt{arXiv:2003.13887}}}.

\bibitem{Maldacena:2016upp}
J.~Maldacena, D.~Stanford, and Z.~Yang, ``{Conformal symmetry and its breaking
  in two dimensional Nearly Anti-de-Sitter space},'' PTEP {\bf 2016} (2016),
  no.~12, 12C104,
\href{http://arXiv.org/abs/1606.01857}{{\texttt{arXiv:1606.01857}}}.

\bibitem{Mertens_2018}
T.~G. Mertens, ``The Schwarzian theory - origins,'' Journal of High Energy
  Physics {\bf 2018} (May, 2018).

\bibitem{Witten:1981mf}
E.~Witten, ``{A Simple Proof of the Positive Energy Theorem},'' Commun.\ Math.\
  Phys. {\bf 80} (1981) 381.

\bibitem{Parker:1981uy}
T.~Parker and C.~H. Taubes, ``{On Witten's Proof of the Positive Energy
  Theorem},'' Commun.\ Math.\ Phys. {\bf 84} (1982) 223.

\bibitem{Wieland:2016dbc}
W.~Wieland, ``{Quasi-local gravitational angular momentum and centre of mass
  from generalised Witten equations},'' Gen. Rel. Grav. {\bf 49} (2017), no.~3,
  38,
\href{http://arXiv.org/abs/1604.07428}{{\texttt{arXiv:1604.07428}}}.

\bibitem{Wieland:2017zkf}
W.~Wieland, ``{New boundary variables for classical and quantum gravity on a
  null surface},'' Class. Quant. Grav. {\bf 34} (2017), no.~21, 215008,
\href{http://arXiv.org/abs/1704.07391}{{\texttt{arXiv:1704.07391}}}.

\end{thebibliography}\endgroup

\end{document}